\documentclass[sigconf]{acmart}
\graphicspath{{figures/}}
\usepackage{calrsfs}
\usepackage{subfigure}

\usepackage{amsmath,amsthm,epsfig,epstopdf}

\DeclareMathAlphabet{\pazocal}{OMS}{zplm}{m}{n}

\usepackage[linesnumbered,ruled,vlined]{algorithm2e}
\usepackage{algpseudocode}
\DeclareMathSymbol{\perp}{0}{symbols}{"3F}
\makeatletter
\def\@copyrightspace{\relax}
\makeatother

\let\emptyset\varnothing
\algnewcommand{\algorithmicor}{\textbf{ or }}
\algnewcommand{\OR}{\algorithmicor}
\AtBeginDocument{%
  \providecommand\BibTeX{{%
    \normalfont B\kern-0.5em{\scshape i\kern-0.25em b}\kern-0.8em\TeX}}}

\setcopyright{none}
\renewcommand\footnotetextcopyrightpermission[1]{} 

\begin{document}

\title{A Load Balanced Recommendation Approach}

\author{Mehdi Afsar, Trafford Crump, and Behrouz Far}

\affiliation{%
 \institution{University of Calgary}
 \city{Calgary}
  \state{AB}
  \country{Canada}
}


\begin{abstract}
Recommender systems (RSs) are software tools and algorithms developed to alleviate the problem of \textit{information overload,} which makes it difficult for a user to make right decisions. Two main paradigms toward the recommendation problem are \textit{collaborative filtering} and \textit{content-based filtering}, which try to recommend the best items using ratings and content available. These methods typically face infamous problems including cold-start, diversity, scalability, and great computational expense.  We argue that the uptake of deep learning and reinforcement learning methods is also questionable due to their computational complexities and uninterpretability.  In this paper, we approach the recommendation problem from a new prospective. We borrow ideas from \textit{cluster head selection} algorithms in \textit{wireless sensor networks} and adapt them to the recommendation problem. In particular, we propose Load Balanced Recommender System (LBRS), which uses a probabilistic scheme for item recommendation.  Furthermore, we factor in the importance of items in the recommendation process, which significantly improves the recommendation accuracy.  We also introduce a method that considers a \textit{heterogeneity} among items, in order to balance the similarity and diversity trade-off.  Finally, we propose a new metric for diversity, which emphasizes the importance of diversity not only from an \textit{intra-list} perspective, but also from a \textit{between-list} point of view. With experiments in a simulation study performed on \textit{RecSim}, we show that LBRS is effective and can outperform baseline methods.
\end{abstract}




\maketitle
\pagestyle{plain} 

\section{Introduction}
\label{sec:intro}
The massive volume of information on the web leads to the problem of  \textit{information overload}, which makes it tough for users to find their items of interest. Many technologies have been developed to help users in this aspect, among which recommender systems (RSs) have been very promising~\cite{adomavicius2005toward, ricci2011introduction}. RSs are software tools and filtering techniques that help users find their items of interest through predicting their ratings on items. Today, the application of RSs is widespread, ranging from e-commerce to news, from e-learning to healthcare~\cite{ricci2011introduction}. 

Traditionally, there are two main methods used in the RS domain: collaborative filtering (CF) and content-based filtering (CBF). The idea behind CF is to find similar users and to generate recommendations based on the assumption that similar users have similar tastes. In contrast, CBF recommends items that are similar to the rating history of the user. These methods typically face severe problems, including cold-start, diversity, scalability, great computational expense, and low quality recommendation~\cite{ricci2011introduction}.  Recent attempts to alleviate part of these problems include deep learning (DL) and deep reinforcement learning (RL) methods.  Nonetheless,  the uptake of DL methods is questionable due to their inherent problems, including data hungriness (this problem is specifically acute regarding the \textit{data sparsity} fact in the RS field~\cite{desrosiers2011comprehensive}), uninterpretability, and high computational complexities~\cite{zhang2019deep}.  While RL is a better option compared to DL (due its abilities in modelling user dynamics, working with implicit feedback~\cite{oard1998implicit}, and taking into account sequential features and long term user engagement~\cite{afsar2021reinforcement}), it also suffers from slow and unstable convergence, uninterpretability, and heavy computation, because the core of deep RL is a DL model~\cite{mnih2015human, lillicrap2015continuous}.

Wireless sensor networks (WSNs) consist of, typically, a huge number of tiny sensor nodes, which are spread within or close to a phenomenon to monitor and collect data~\cite{akyildiz2002wireless}. Although the possibility of adopting energy harvesting technologies has been investigated for these networks~\cite{afsar2019load}, power efficiency is still the most important challenge in these networks, due to the limited computational power and capacity of sensor nodes.  Today, WSNs is a building block in the internet of things (IoT) technology and are applied in many applications~\cite{mainetti2011evolution, kandris2020applications}. {\it Clustering} is an effective method in WSNs to make them more energy efficient~\cite{abbasi2007survey}, which divides the network architecture into two layers: cluster heads and cluster members. Typically, cluster members sense their surroundings and send their data to cluster heads, which are responsible to aggregate this data and transmit them to a remote station, known as \textit{base station}.  The heavy and important task of being a cluster head makes \textit{cluster head selection} the main problem in clustered WSNs, as cluster heads should usually be rich in some features, like remaining energy, compared to other nodes~\cite{afsar2014clustering, shahraki2020clustering}. 

With carefully studying both problems, namely item selection for recommendation in RSs and cluster head selection in WSNs, we recognized that they are analogous. In particular, if we assume that items in RSs are similar to sensor nodes in WSNs, items features, e.g., quality, are similar to nodes features, like remaining energy, and high quality items are similar to cluster head candidates, then we can assume that item selection for recommendation is analogous to cluster head selection in WSNs. This analogy motivated us to prepare this paper and to investigate the possibility of applying methods from clustering in WSNs to the recommendation domain. Therefore, we approach the recommendation problem from a new perspective and aim at solving the problems of existing RS methods---i.e., cold-start, diversity, scalability, great computational expense, and uninterpretability---using the idea of \textit{load balancing} in clustered WSNs~\cite{heinzelman2000energy}. The key idea behind load balancing is to rotate the heavy task of being cluster head among all sensor nodes periodically, so as the load is fairly distributed among all nodes and the network longevity is improved. We adapt this idea to the recommendation problem.

More precisely, with the idea of load balancing in mind, we propose three variants of Load Balanced Recommender System (LBRS). Similar to~\cite{heinzelman2000energy}, basic LBRS (B-LBRS) uses a probabilistic scheme for item recommendation, which can maximize the recommendation diversity.  To improve recommendation accuracy, we extend B-LBRS to factor in the importance of items in the recommendation strategy and call it priority LBRS (P-LBRS). 
To maximize user engagement, a good RS should provide a balanced level of accuracy and diversity in recommendations (known as similarity and diversity trade-off~\cite{bradley2001improving}), something that P-LBRS fails to provide. Therefore, we present H-LBRS, which considers a {\it heterogeneity} among items and can provide a flexible framework to balance the similarity and diversity trade-off. Not only the proposed methods are fast and explainable, they can solve the cold-start problem and scale well to the number of users and items.  We study the performance of the proposed methods using a simulation study performed on RecSim~\cite{ie2019recsim}, an RS simulator developed by Google, and the results of experiments are promising. In general, our contributions are as follows:

\begin{enumerate}
\item We approach the recommendation problem from a totally new perspective and apply methods from cluster head selection problem in clustered WSNs. To the best of our knowledge, no one has examined this idea before.

\item We take one step further and incorporate items importance into recommendation, which significantly improves the recommendation accuracy.

\item We also extend LBRS to incorporate the heterogeneity of items into the recommendation mechanism, which aims at balancing the similarity and diversity trade-off.

\item A novel diversity metric is introduced, which takes into account the importance of \textit{between-list} diversity.

\item Using a simulation study performed on RecSim, we validate the performance of the proposed LBRS.

\end{enumerate}

This paper is organized as follows. Section~\ref{sec:back} provides a quick background to introduce the topics discussed in the paper.  Section~\ref{sec:method} explains the proposed method in detail. Section~\ref{sec:exp} presents the results of experiments and the paper is finally concluded in section~\ref{sec:con}.

\section{Background}
\label{sec:back}
In this section, we introduce the main concepts used in the paper. We first provide a background on RSs and its popular techniques. Then, we explain LEACH~\cite{heinzelman2000energy}, a popular clustering algorithm in WSNs on which our work is based.

\subsection{Recommender Systems}
\label{subsec:RS}
CF has been the foundation of first RSs developed~\cite{ricci2011introduction}. The idea was simple: people tend to rely on recommendations made by their peers~\cite{ricci2011introduction}.  In fact, this was the rationale behind the first RS, Tapestry~\cite{goldberg1992using}, and they termed it as \textit{collaborative filtering}.  Later, this term was broadened to \textit{recommender systems} to reflect two facts~\cite{resnick1997recommender}: 1) the method may not collaborate with users, 2) the method may suggest interesting items, not filter them. Another approach toward the recommendation problem is CBF, where the idea is to recommend items similar to the user \textit{profile}, which is a structured representation of user interests~\cite{pazzani2007content, lops2011content}.  These methods, i.e., CF and CBF, fail to provide effective recommendations due to their problems stated above---cold-start, diversity, scalability, great computational expense, and low quality recommendation. 
Not long ago, unprecedented successes in the machine vision field encouraged RS researchers to employ DL methods in the RS field~\cite{van2013deep, wang2015collaborative, covington2016deep, cheng2016wide}. However, as mentioned earlier, DL models are data hungry, incomprehensible or unexplainable, and require massive computational resources~\cite{zhang2019deep}.  

Recently, many recommendation techniques have been proposed using deep RL methods~\cite{zhao2017deep, zhao2018deep, zhao2018recommendations, chen2018stabilizing, zou2019reinforcement}. Applying RL to RSs, with huge action and state spaces, was not possible until the advent of deep Q network (DQN)~\cite{mnih2015human}, which is the combination of traditional Q-learning algorithm~\cite{watkins1992q} and convolutional neural networks.
The use of RL in RSs has several advantages compared to the previous methods, including CF, CBF, and DL; first and foremost, RL is an effective method to model the sequential and dynamic nature of interaction between users and the RS. Second, RL can effectively work with implicit feedback~\cite{oard1998implicit}. Finally, unlike \textit{myopic} methods (e.g., CF, CBF, and DL), it can maximize long term user engagement.  Nevertheless, deep RL methods also suffer from problems similar to DL models, including slow convergence, unexplainability, and great computational expense. In section~\ref{sec:exp}, we compare the performance of LBRS with a DQN-based RS.  

In contrast to the previous work, the proposed LBRS is simple, fast, scalable, requires no rating from users, and can recommend a diverse range of items.  To the best of our knowledge, we are the first to propose a recommendation approach using ideas from clustered WSNs.

\subsection{LEACH}
Low-energy adaptive clustering hierarchy (LEACH)~\cite{heinzelman2000energy, heinzelman2002application} is a popular clustering algorithm developed for WSNs. As stated earlier, the main idea in LEACH is to ensure that all nodes in the network play the role of cluster head once in a predefined time, which is known as load balancing in the WSN literature~\cite{abbasi2007survey, afsar2014clustering}. In LEACH, the network operational time is assumed to be divided into \textit{rounds} and each round has two phases: \textit{setup} and \textit{steady-state}.  In the setup phase, clusters are formed and in the steady-state phase, the data is transmitted to the base station.  LEACH uses a probabilistic approach for cluster head selection. More precisely, it is assumed that all nodes are synchronized and can start the network operations at the same time.  At the beginning of each round, all the nodes compute a threshold as:
\begin{equation}
T(i) = 
    \begin{cases}
      \frac{p}{1-p \cdot ~(r ~\textrm{mod}~\frac{1}{p})} & \text{if $i \in G$}\\
      0 & \text{otherwise,}
    \end{cases} 
    \label{eq:leach}
\end{equation}
where $G$ is the set of nodes that have not been cluster head in the last $\frac{1}{p}$ rounds, $p$ is the optimal probability of cluster head selection, and $r$ is the current round. Then, they generate a random number between 0 and 1. If the random number is less than $T(i)$, node $i$ selects itself as a cluster head in round $r$ and advertises its decision within its surrounding to form its cluster. Otherwise, the node remains as a regular node and joins the nearest cluster head. This way, it is guaranteed that all the nodes are selected as cluster head once in $\frac{1}{p}$ rounds.   After this duration, they are eligible to be selected as the cluster head again.


\section{Methodology}
\label{sec:method}

\subsection{Problem Definition}
\label{subsec:PD}
We study the recommendation problem in which an RS, in each time step $t$, recommends a list of items $L=(i_1, i_2, ..., i_k)$ to a user $u \in \pazocal{U}$, where $i \in \pazocal{I}$, $k \in \mathbb{Z}$, $N=|\pazocal{U}|$, and $M=|\pazocal{I}|$.  Each user $u$ interacts with the RS in a recommendation \textit{session}, which composes of several time steps $(t_1, t_2, ..., t_{\tau})$, where~finite~$\tau\in\mathbb{Z}$. Note that the user selects either one item or none from $L$. The length of each session depends on the \textit{time budget} ($B_t$) of the user. We assume that when $B_t$ finishes, the user leaves the system and never comes back\footnote{This assumption is based on the user behavior model discussed in~\cite{craswell2008experimental}.}. Then, the RS can start a new recommendation session with a new user.  Similar to RL literature, it is assumed that each user provides a numerical reward, indicated by $R$, on selecting one item from $L$. In fact, $R$ reflects the \textit{similarity} between user interest and recommended item. On the other hand, a good recommendation approach should provide a diverse range of items~\cite{bradley2001improving}. This diversity could be in terms of topic, quality, etc. Our RS objective is to recommend a list of items to the user in each time step $t$, which not only maximizes $R$, but it can also provide a proper level of diversity. 

\subsection{The Proposed LBRS}
\label{subsec:lbrs}

\subsubsection{Basic LBRS (B-LBRS)}

Similar to LEACH, B-LBRS uses a probabilistic approach for item recommendation.  However, unlike LEACH in which cluster head selection is distributed and each node individually comes to a decision on whether select itself to be a cluster head, the recommendation process in our system is a centralized scheme performed by the RS agent. Therefore, we can rewrite Eq.~\eqref{eq:leach} as 
\begin{equation}
T_b(i) = 
    \begin{cases}
      \frac{p}{1-p\cdot (t ~\textrm{mod}~\frac{1}{p})} & \text{if $i \in \pazocal{G}$}\\
      0 & \text{otherwise,}
    \end{cases} 
    \label{eq:basic}
\end{equation}
where $T_b(i)$ is the threshold for item $i$ to be recommended at $t$ and $p$ is the probability of recommending items. If $i$ has been recommended to the user in the last $\frac{1}{p}$ steps (i.e., $i \notin \pazocal{G}$), the probability of being recommended at $t$ is 0.  Otherwise, the RS agent calculates $T_b(i)$ according to the upper case of Eq.~\eqref{eq:basic} and recommends $i$ to the user if the generated random number is less than $T_b(i)$.

Eq.~\eqref{eq:basic} guarantees that all the items are recommended to users once in $\frac{1}{p}$ time steps, but it fails to control the number of items to recommend in each time step (i.e., $k$). For example, with a small $M=100$ and $p=0.05$ (a popular value in WSN literature), all the items are recommended to the user in 20 time steps and average $k=5$. However, we observed in experiments that $k$ could be very noisy, with as large as 20 or 30 in one time step and as small as one in another time step. This is not definitely acceptable in a sensitive field like RSs whose users are humans. Instead, we use a deterministic approach in LBRS and the element that defines the number of items to recommend in each time step is a predefined $k$. While there is no optimal value for $k$, there are some studies that examine the effect of this number on user satisfaction and suggest that a $k\in[5,10]$ could be a good choice~\cite{bollen2010understanding, beierle2019choice}.  We simply let $p=\frac{k}{100}$ and the maximum number of items to recommend in one time step is $k$.  We empirically discuss this in section~\ref{sec:exp}.


\subsubsection{Priority LBRS (P-LBRS)}

A problem with LEACH in the WSN literature is that it does not factor in the importance of nodes in cluster head selection and selects them randomly, while some nodes may have higher remaining energy than others and it is better to select them as cluster heads. Similarly, in our recommendation problem, some items may be of higher quality or are more important than others and it is wise to recommend them to the user more often. Accordingly, we incorporate items importance into B-LBRS and call it P-LBRS. The most straightforward way to do this is to directly factor in the quality of items into Eq.~\eqref{eq:basic} as
\begin{equation}
T_{p}(i) = 
    \begin{cases}
      \Big[\frac{p}{1-p \cdot (t ~\textrm{mod}~\frac{1}{p})}\Big]\cdot Q(i) & \text{if $i$ in $\pazocal{G}$} \\
      0 & \text{otherwise,}
    \end{cases} 
    \label{eq:p-lbrs1}
\end{equation}
where $Q$ is the quality of item $i$ and is defined by the designer. As it is possible that $Q(i) < 0$ (see section~\ref{sec:exp}), a problem with Eq.~\eqref{eq:p-lbrs1} is that $T_p(i)$ could be negative. Moreover, for $Q(i)>1$, it is possible that $T_p(i)>1$. To fix these problems, we first normalize $Q(i)$ using
\begin{equation}
Q_{\textrm{norm}}(i) = \frac{Q(i) - Q_{\textrm{min}}}{Q_{\textrm{max}} - Q_{\textrm{min}}},
\label{eq:norm}
\end{equation}
where $Q_{\textrm{min}}$ and $Q_{\textrm{max}}$ are the minimum and maximum quality of items, respectively, and are defined by the designer (more details in section~\ref{sec:exp}). Then, Eq.~\eqref{eq:p-lbrs1} can be rewritten as 
\begin{equation}
T_{p}(i) = 
    \begin{cases}
      \Big[\frac{p}{1-p\cdot (t ~\textrm{mod}~\frac{1}{p})}\Big]\cdot Q_{\textrm{norm}}(i) & \text{if $i$ in $\pazocal{G}$} \\
      0 & \text{otherwise.}
    \end{cases} 
    \label{eq:p-lbrs2}
\end{equation}
Eq.~\eqref{eq:p-lbrs2} guarantees that items with a higher quality are more likely to be selected for recommendation. 

\subsubsection{Heterogeneous LBRS (H-LBRS)}
One problem with P-LBRS is that it might perform greedily and only high quality items get selected for recommendation. This negatively influences the diversity, and consequently, the satisfaction of the user. To solve this problem, we introduce H-LBRS. Inspired by~\cite{smaragdakis2004sep}, H-LBRS assumes that items in $\pazocal{I}$  are \textit{heterogeneous} and divides them into \textit{high-quality} items $\pazocal{I}_{h} \subset \pazocal{I}$ and \textit{low-quality} items $\pazocal{I}_{l} \subset \pazocal{I}$, where $\pazocal{I}_{h} \cup \pazocal{I}_{l} = \pazocal{I}$ and $\forall i\in \pazocal{I}_{h}$, $\forall j \in \pazocal{I}_{l}$, $Q(i) > Q(j)$.  Accordingly, there are two thresholds, $T_{h}$ and $T_l$, for item selection at $t$, where $T_{h}$ incorporates the probability of high-quality items ($p_{h}$) into item recommendation and $T_l$ factors in the probability of low-quality items ($p_l$). These probabilities are calculated as
\begin{equation}
p_{h}=\frac{p}{1 + \lambda \cdot f} \cdot (1+ \lambda),
\label{eq:ph}
\end{equation}
and
\begin{equation}
p_{l}=\frac{p}{1 + \lambda \cdot f},
\label{eq:pl}
\end{equation}
where $\lambda$ is the coefficient of heterogeneity and $f$ is the fraction of $\pazocal{I}_h$.  For example, if  $f=0.1$ and $\lambda=2$,  $|\pazocal{I}_h|=\frac{M}{10}$ and $\forall i \in \pazocal{I}_h$ has two times more priority than $\forall j \in \pazocal{I}_l$. Consequently, there are two thresholds for item selection:
\begin{equation}
T_{h}(i) = 
    \begin{cases}
      \frac{p_{h}}{1-p_{h}\cdot (t ~\textrm{mod}\frac{1}{p_{h}})} & \text{if $i$ in $\pazocal{G}_h$}\\
      0 & \text{otherwise,}
    \end{cases} 
    \label{eq:hlbrs-h} 
\end{equation}
and
\begin{equation}
T_l(j) = 
    \begin{cases}
      \frac{p_{l}}{1-p_{l}\cdot (t~ \textrm{mod}\frac{1}{p_{l}})} & \text{if $j$ in $\pazocal{G}_l$}\\
      0 & \text{otherwise.}
    \end{cases}
    \label{eq:hlbrs-l} 
\end{equation}
H-LBRS provides a flexible framework to select the level of diversity in recommendations.  There are two parameters $\lambda$ and $f$ that affect the performance of H-LBRS. The selection of $\lambda$ depends on the application and the level of diversity required. Moreover, we control $f$ by defining a quality threshold, $Q_{\mathrm{th}}$.  In other words, $i\in \pazocal{I}_h \Leftrightarrow Q(i) \geq Q_{\mathrm{th}}$.  In the next section, we empirically show the effect of $\lambda$ and $Q_{\mathrm{th}}$ parameters on the performance of H-LBRS.   Moreover, Algorithm~1 demonstrates the pseudo code of item recommendation in LBRS---for the sake of brevity, it demonstrates  the three versions of LBRS).

\begin{algorithm}
\caption{Item Recommendation in LBRS}
\SetAlgoLined
\DontPrintSemicolon

\textbf{initialization} ($k = 5$, $p=\frac{k}{100}$, $\pazocal{G}/\pazocal{G}_l/\pazocal{G}_h \leftarrow \emptyset$, $L \leftarrow \emptyset$)\\
\For {\normalfont{\textbf{all}} $u  \in \pazocal{U} $}
{
\While {$ B_t(u) \geq \ell_d $}
{
\While {$ |L| \leq k $}
{
\For {\normalfont{\textbf{all}} $i  \in \pazocal{I} $}
{
\If{$i \in \pazocal{G} / \pazocal{G}_h/\pazocal{G}_l$}
{
calculate $T_b(i)$/$T_p(i)$/$T_h(i)$/$T_l(i)$ using Eqs.~\eqref{eq:basic}/\eqref{eq:p-lbrs2}/\eqref{eq:hlbrs-h}/\eqref{eq:hlbrs-l}\\
\If{\normalfont{rand(0, 1)} < $T_b(i)$/$T_p(i)$/$T_h(i)$/$T_l(i)$}{add $i$ to $L$ and $\pazocal{G}/ \pazocal{G}_h/\pazocal{G}_l$}
}
}
}
update $B_t(u)$\\
calculate $p_h$/$p_l$ using Eqs.~\eqref{eq:ph}/\eqref{eq:pl}\\
\If{($t$ \normalfont{mod} $\frac{1}{p}/\frac{1}{p_l}/\frac{1}{p_h} = 0$)}{$\pazocal{G}/\pazocal{G}_h/\pazocal{G}_l \leftarrow \emptyset$}
$L \leftarrow \emptyset$\\
}
}

\label{alg:lbrs}
\end{algorithm}

\section{Experiments}
\label{sec:exp}
In this section, we present the experiments conducted to show the performance of LBRS. We use RecSim~\cite{ie2019recsim} to examine the performance of the proposed method.  It is noteworthy to mention that all recommendation methods are implemented in Python and experiments are conducted on AAA cluster at the University of BBB.

\subsection{Setup}
RecSim is a configurable simulation platform for RSs developed by Google. It provides RS researchers with the opportunity to develop various models of users and RS dynamics.  It is noteworthy to mention that while LBRS is a generic RS, we examine its performance in a document recommendation use case, similar to~\cite{ie2019reinforcement}.  Except for the differences mentioned below, we configure RecSim exactly similar to the simulation study conducted in~\cite{ie2019reinforcement} (described in section 6 of that paper). The differences between our setup and theirs are as follows.  First, we assume that there are $M$ documents in the local database of our RS to recommend to the user. This is in contrast with the setting described in~\cite{ie2019reinforcement}, in which they assume that at each $t$, $m$ candidate documents are drawn from a large corpus and then a list of size $k$ from these $m$ candidates are recommended to the user.  Second, unlike~\cite{ie2019reinforcement}, user interest is not observable to the RS in our setting, which is the case in the real.  Third, we assume that with probability 0.5 all users select no item from the list recommended across time steps (explained shortly).  Other settings of the experiments and assumptions are exactly the same as that in~\cite{ie2019reinforcement}. For completeness, we briefly describe these settings below. 

For statistical significance, there are $N=5000$ users interacting with the RS and $M=10$,000 documents in $\pazocal{I}$ for recommendation.  Documents are categorized into $T=20$ topics and each document $d_i\in \pazocal{I}$ is associated with only one topic. While it is possible to associate each document to multiple topics~\cite{zou2019reinforcement}, we will not study this case in this paper. For simplicity, it is assumed that the length of all documents are similar and constant.  Each document has an \textit{inherent quality} ($Q$), which shows the attractiveness of the document to the average user. The quality is randomly distributed across documents. The topics are divided into high quality and low quality topics. Of $T$ topics, $\lfloor \frac{1}{3} \rfloor$ of topics are of high quality and $\forall d_i \in \pazocal{I}_h$, $Q(d_i)\in [0, Q_{\textrm{max}}]$, where $Q_{\textrm{max}}=3$. The remaining topics are of low quality and $\forall d_j \in \pazocal{I}_l$, $Q(d_j)\in [-Q_{\textrm{max}}, 0]$. Users interest ($I_T$) to topics is modelled as a real number $\in [-1, 1]$,  where -1 is totally uninterested and 1 is fully interested. Thus, $I_{T}\in \mathbb{R}^{T}$ is a vector with length $T$ and each element shows user interest to a specific topic ($I_t$).  It is possible that $I_t$ changes after consuming a document from a specific topic. This change is computed as $\delta = (-y |I_t|+y)\cdot -I_t$, where $y \in [0,1]$ ($y=0.3$ adopted in experiments). 	A positive change in user interest, i.e., $I_t = I_t + \delta$, is occurred with probability $\frac{(I_t+1)}{2}$ and a negative change, i.e., $I_t = I_t - \delta$, with $\frac{(I_t-1)}{2}$.  The document utility for a user is modelled as $S(u, d) = (1-\gamma)I_t(d) + \gamma Q(d)$. Each document consumed by the user has the potential to replenish the user budget $B_t$. This is called \textit{bonus} and is calculated as $b = (0.9/3.4) \cdot \ell \cdot S(u,d)$, where $\ell$ indicates how many time steps it takes to consume a document ($\ell=4$ in experiments).  It is possible for a user to not select a document from $L$. This choice is called \textit{null} item and is the $(k+1)th$ item in $L$. For simplicity, we consider a fixed probability $P(\perp|L)=0.5$ for all users to select the null item across time steps. Selecting a null item also consumes $\ell=1$ unit from $B_t$. User behavior in selecting a document from $L$ is modelled using \textit{user choice} model. We use the general \textit{conditional choice model} defined in~\cite{ie2019reinforcement}.  For a complete and more detailed description of these settings and assumptions, refer to~\cite{ie2019reinforcement}.

As stated in section~\ref{subsec:PD}, the user provides a numerical reward $R$ upon receiving the recommendation. We assume a fixed $R=4$ when the user selects a document from $L$. Otherwise, $R=0$. Table~\ref{tab:param} presents the value of parameters used.

\begin{table}
\caption{Parameters Values}
\centering
\small
\begin{tabular}{| c | c |}
\hline
{\bf Parameter} & {\bf Value} \\ [0.5ex]
\hline
$N$ & 5000 \\
\hline
$M$ & 10,000 \\
\hline
$k$ & 5\\
\hline
$Q_{\mathrm{max}}$ & 3\\
\hline
$T$ & 20 \\
\hline
$y$ & 0.3\\
\hline
$\alpha$ & 1 \\
\hline
$\beta$ & 1 \\
\hline
$\gamma$ & 1 \\
\hline
$B_t$ & 200\\
\hline
$\ell_d$ & 4\\
\hline
$\ell_\perp$ & 1\\
\hline
$P(\perp |L)$ & 0.5\\
\hline
$R$ & 4\\
\hline

\end{tabular}
\label{tab:param}
\end{table}

In order to compare the performance of the proposed LBRS with other works, we have selected the following baselines. Since we are using RecSim to evaluate the performance, we should use methods that only work with implicit feedback. All CF, CBF, and DL methods are not suitable to use as the baseline, since there is no rating and content  available to apply these methods. Accordingly, we have implemented the following baselines:
\begin{itemize}

\item \textbf{DQN.}  We have implemented a DQN~\cite{mnih2015human} agent with the following specifications. Each state $s\in S$ encodes information about the history of the user (last $m$ documents selected by the user). In addition to documents indices, documents topics and qualities are encoded in the states. No context from user, but user index, is used in state encoding.  Each action $ a \in A$ is to recommend a list ($L$) of $k$ documents to the user.  We use a \textit{$k$ nearest neighbor} method for our DQN agent to generate lists of size $k$ as actions\footnote{In SlateQ~\cite{ie2019reinforcement}, each list $L$ is an action and the action space $|A|={N \choose k}$. While this setting is possible in a two-stage RS in which the first stage narrows down $M$ to $m$ candidate items, where $M>>m$, it is not simply scalable in our use case as with a not very large $M=1000$, we have $|A|={1000 \choose 5}\simeq 8\times10^{12}$.}. The same neural network architecture as that described in~\cite{mnih2015human} is used for the Q network. Other DQN parameters are listed in Table~\ref{tab:dqn-param}.

\begin{table}
\caption{DQN Hyper-parameters Values}
\centering
\small
\begin{tabular}{| l | c |}
\hline
{\bf Hyper-parameter} & {\bf Value} \\ [0.5ex]
\hline
Discount factor & 0.99 \\
\hline
$m$ & 10 \\
\hline
Memory size & 1M transitions \\
\hline
Minibatch size & 32 \\
\hline
Min history to start learning & 1000 transitions\\
\hline
Decayed $\epsilon$ & 1$\rightarrow$ 0.1\\
\hline
Optimizer & Adam\\
\hline
Learning rate & 0.1\\
\hline
Loss & Mean-squared error\\
\hline

\end{tabular}
\label{tab:dqn-param}
\end{table}

\item \textbf{$\epsilon$-greedy.} We have also implemented an $\epsilon$-greedy agent with $\epsilon=0.1$, which is a popular method in \textit{multi-armed bandits}~\cite{sutton1998reinforcement}. $\epsilon$-greedy explores with the probability of $\epsilon$ and exploits with $1-\epsilon$.

\item \textbf{Random.} This agent randomly selects lists of documents from $\pazocal{I}$.
\end{itemize}

 In the experiments conducted, we use the following performance metrics: 
 
 \begin{itemize}
 
 \item  \textbf{Average reward.} This metric, indicated by $\pazocal{R}$, measures the similarity between users interests and documents recommended and is calculated as:
 
\begin{equation}
\label{eq:avg-rew}
\pazocal{R}=\frac{\sum_{u=1}^N \sum_{t=1}^{\tau_u} R_t(u)}{\sum_{u=1}^N \sum_{t=1}^{\tau_u} 1}.
\end{equation}
 
 \item  \textbf{Average diversity:} This metric, indicated by $\pazocal{D}$, measures the variety between recommended items. Two parameters are used for diversity metric. The first one measures \textit{intra-list similarity (ILS)}  and is calculated as~\cite{ziegler2005improving}:
 
\begin{equation}
ILS (L)=\frac{\sum_{i \in L} \sum_{j \in L, i \neq j}  Sim(i, j)}{\sum_{i \in L} \sum_{j \in L, i \neq j} 1},
\label{eq:ils}
\end{equation}
where $Sim(i, j)$ is the cosine similarity between items $i$ and $j$. To check the similarity between documents in our use case, we consider two features for each document: document topic and document quality.  For document quality, two documents $d_i$  and $d_j$ are considered similar if $Q(d_i)>0$ AND $Q(d_j) > 0$ or $Q(d_i)<0$ AND $Q(d_j)< 0$. In other words, $d_i$ and $d_j$ are similar if $d_i, d_j \in \pazocal{I}_h$ or $d_i, d_j \in \pazocal{I}_l$.
This measure (i.e., $ILS$) only checks the similarity within a list of items; however, in a sequential interaction between the user and the RS, it is important to track the similarity between consecutive lists recommended. Accordingly, we introduce a novel parameter, called \textit{between-list similarity (BLS)}, and is calculated as
 \begin{equation}
BLS (L_n, L_{n+1})=\frac{\sum_{i \in L_n} \sum_{j \in L_{n+1}}  \pazocal{S}(i, j)}{\sum_{i \in L_n} \sum_{j \in L_{n+1}} C},
\label{eq:bls}
\end{equation}
where  $\pazocal{S}(i, j)=Sim(i, j)$ and $C=1$ if $i \neq j$. Otherwise,  $\pazocal{S}(i, j)=k$ and $C=k$.  These values have been adopted to specifically penalize the RS method if an item is repeated across lists of recommendations. $\pazocal{D}$ is then calculated as 
 \begin{equation}
\pazocal{D} = \frac{\alpha \cdot ILS + \beta \cdot BLS}{2},
\label{eq:div}
\end{equation}
 where $\alpha$ and $\beta$ are weights to define the importance of each diversity parameter, namely $ILS$ and $BLS$. For convenience, we let $\alpha=\beta=1$. Finally, similar to Eq.~\ref{eq:avg-rew}, the average diversity is computed as 
 \begin{equation}
\label{eq:avg-div}
\overline{\pazocal{D}}=\frac{\sum_{u=1}^N \sum_{t=1}^{\tau_u} \pazocal{D}_t}{\sum_{u=1}^N \sum_{t=1}^{\tau_u} 1}.
\end{equation}
 While we report the results of diversity according to Eq.~\eqref{eq:avg-div} in the next section, we indicate the average diversity by simple $\pazocal{D}$ for convenience.
 
\end{itemize}

\subsection{Results}
\label{subsec:res}

In this section, the results of the simulation study conducted on RecSim are presented. All results are statically significant and within a $95\%$ confidence interval.

\subsubsection{Parameter Study}
\begin{figure}
\includegraphics[width=\linewidth]{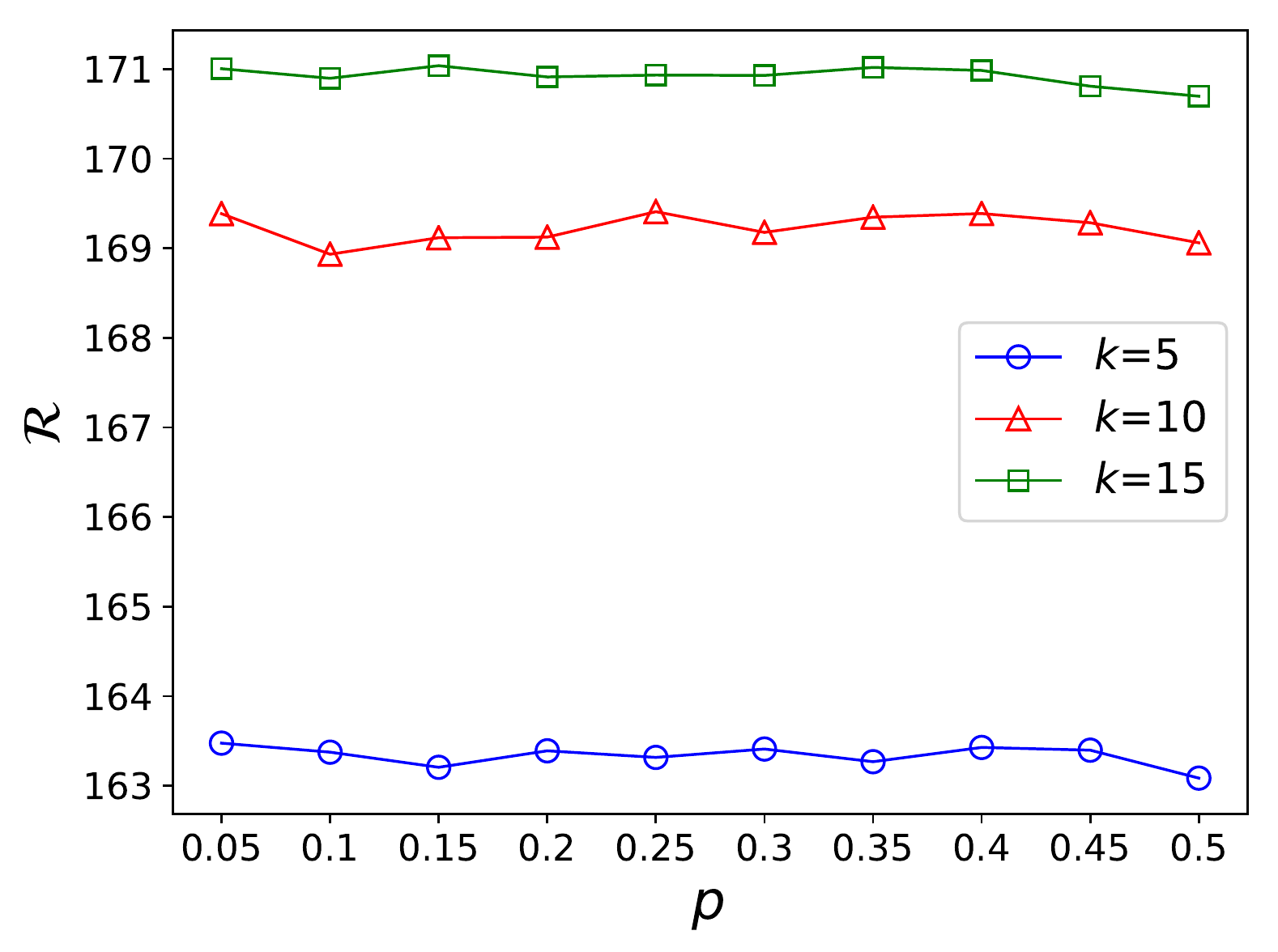}
\caption{The effect of $p$ on $\pazocal{R}$ in B-LBRS }
\label{fig:blbrs-R}
\end{figure}

\begin{figure}
\includegraphics[width=\linewidth]{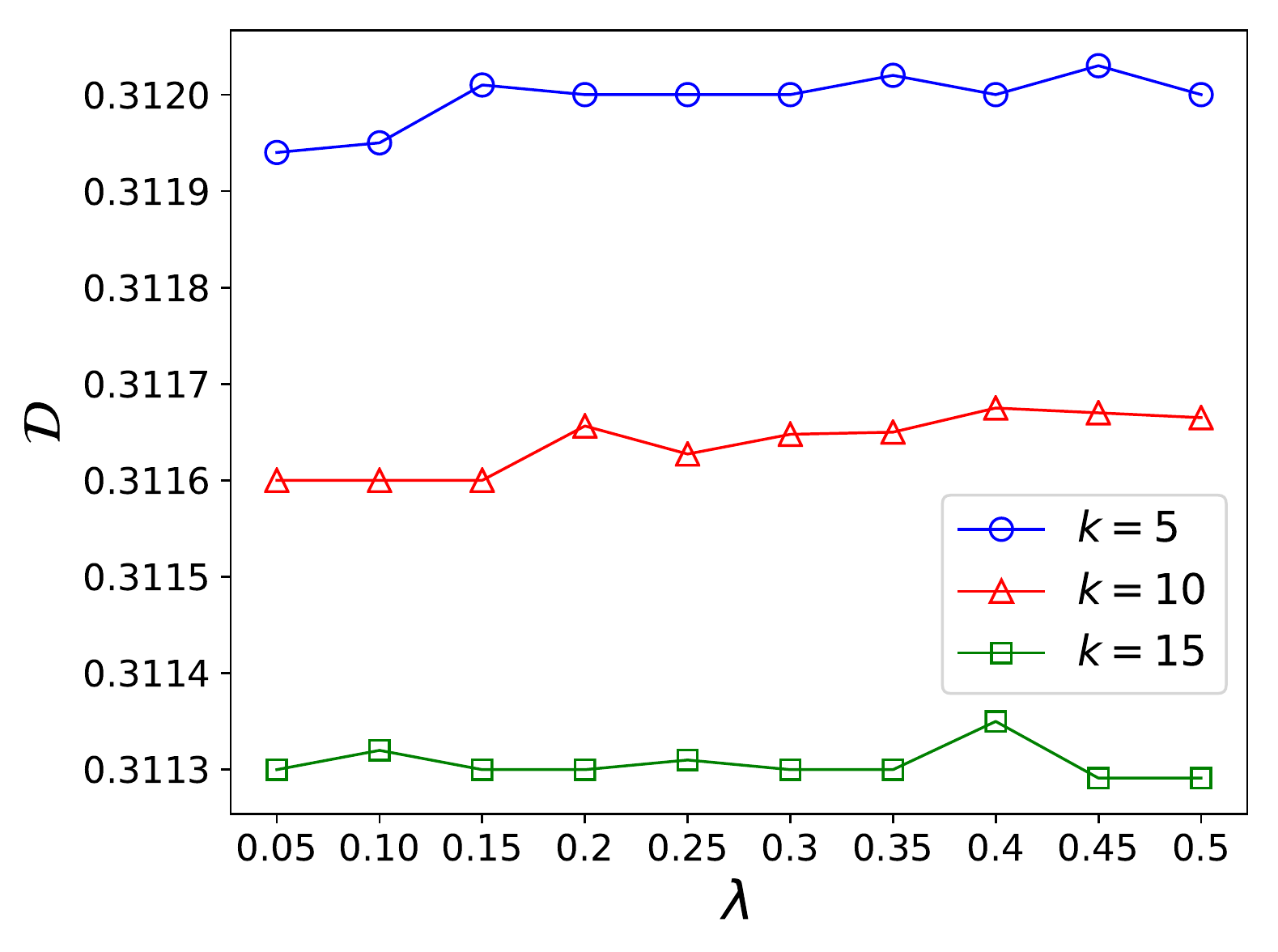}
\caption{The effect of $p$ on $\pazocal{D}$ in B-LBRS}
\label{fig:blbrs-Div}
\end{figure}

We first study the effect of different parameters on the performance of the three methods B-LBRS, P-LBRS, and H-LBRS.   Figs.~\ref{fig:blbrs-R} and~\ref{fig:blbrs-Div} depict performance metrics for B-LBRS. In general, it is observable from figures that when $k$ controls the number of documents recommended to the user, the effect of $p$ on $\pazocal{R}$ and $\pazocal{D}$ is almost negligible.  Moreover, when $k$ increases, both $\pazocal{R}$ and $\pazocal{D}$ are improved, although this improvement is more tangible in $\pazocal{R}$. In particular, the improvement in $\pazocal{R}$ is around $5\%$ with $k$ changing from 5 to 15, while this number is around $0.2\%$ for $\pazocal{D}$.  The reason of this improvement for $\pazocal{R}$ is perhaps in light of the larger $k$, the more likely that the documents of interest to the user to be among recommended items.  Also, when $k$ is larger, the probability of having a more diverse list grows, as more items from different topics and qualities could be among the recommended items. It is worthy to note that, in Fig.~\ref{fig:blbrs-Div}, a smaller $\pazocal{D}$ score means a better diversity. 

\begin{figure}
\includegraphics[width=\linewidth]{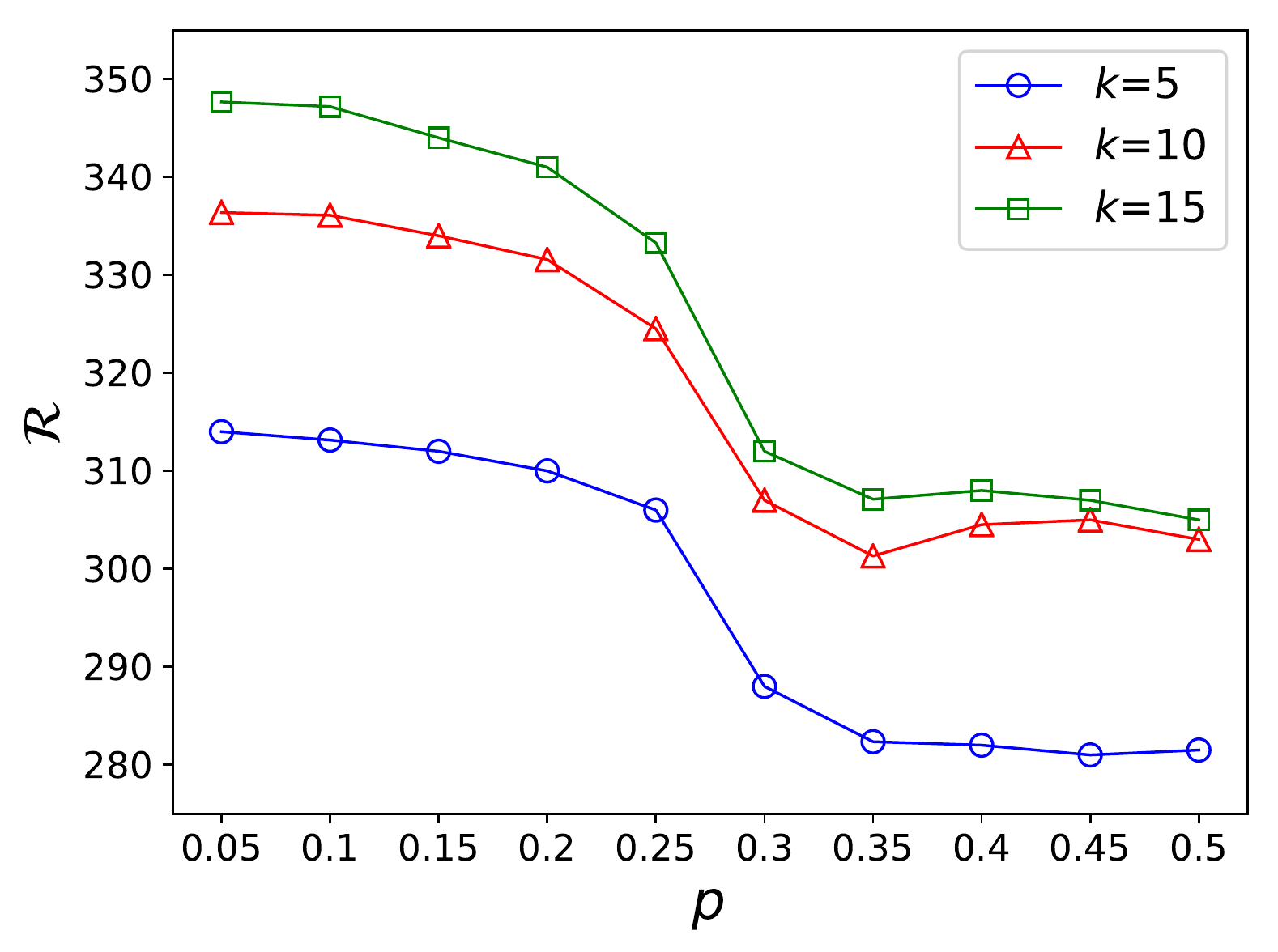}
\caption{The effect of $p$ on $\pazocal{R}$ in P-LBRS}
\label{fig:plbrs-R}
\end{figure}

\begin{figure}
\includegraphics[width=\linewidth]{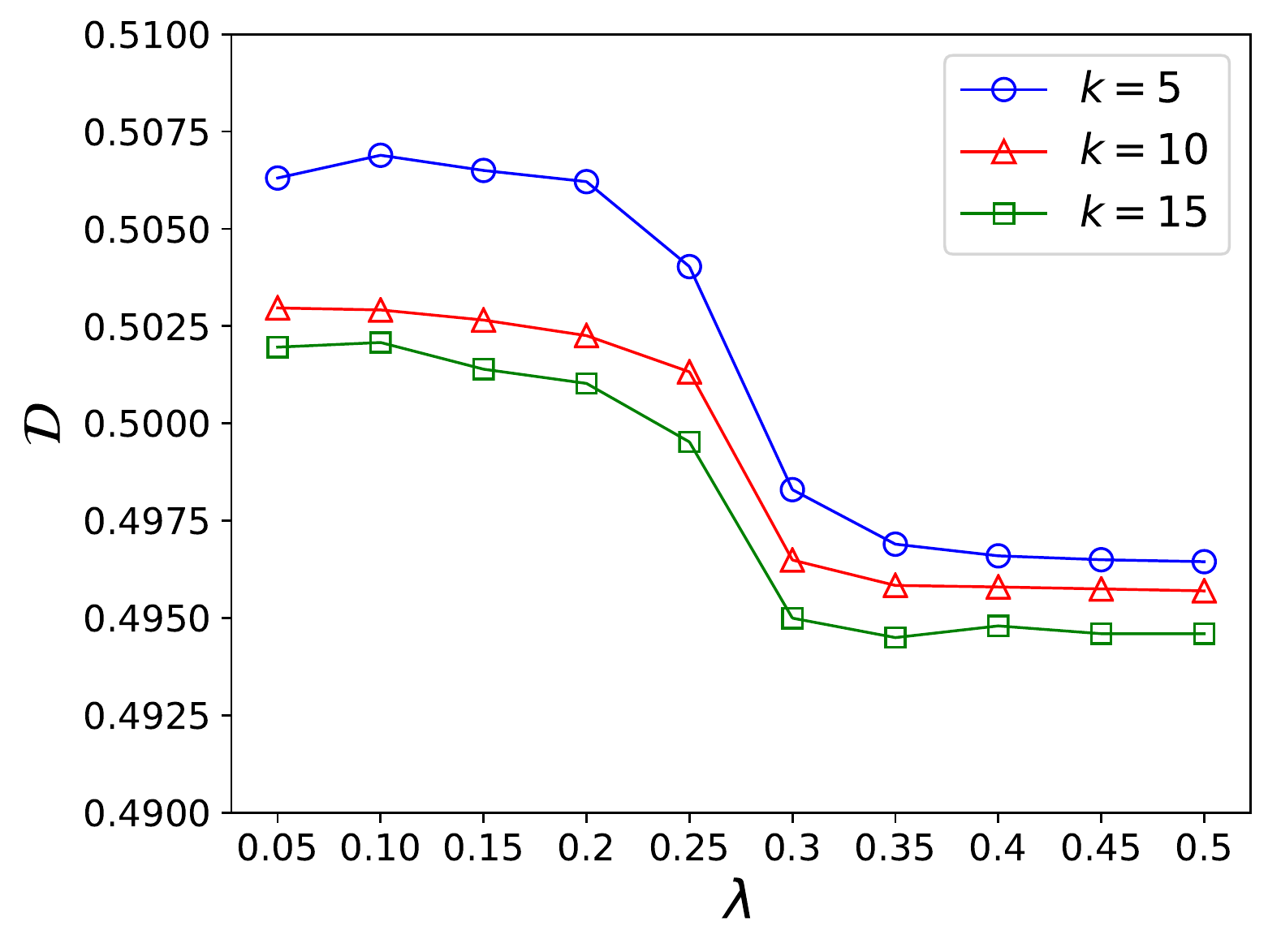}
\caption{The effect of $p$ on $\pazocal{D}$ in P-LBRS}
\label{fig:plbrs-Div}
\end{figure}

On the contrary, Figs.~\ref{fig:plbrs-R} and~\ref{fig:plbrs-Div} illustrate that P-LBRS is sensitive to the value of $p$, such that $\pazocal{R}$ and $\pazocal{D}$ experience a steady plummet with the increase in the value of $p$.  A reason for drop in $\pazocal{R}$ is perhaps the fact that when $p$ is larger, according to Eq.~\eqref{eq:p-lbrs2}, 
$T_p$ is generally larger so as documents with lower qualities also have a good chance to get selected for recommendation.  Also, a larger $p$ shrinks the $\frac{1}{p}$ duration, again increasing the chance for all documents with any quality to be eligible for recommendation. This affects both user interest and budget, leading to a diminished $\pazocal{R}$.  Fig.~\ref{fig:plbrs-Div} demonstrates that a larger $p$ can improve diversity, probably with the same justification mentioned for $\pazocal{R}$, i.e., a more diverse set of nodes have the chance to get selected for recommendation. Also, similar to B-LBRS, a large $k$ can improve both $\pazocal{R}$ and $\pazocal{D}$.  Compared to B-LBRS, while $\pazocal{R}$ is improved by almost 100$\%$ with P-LBRS, this improvement is at the cost of sacrificing $\pazocal{D}$, which diminishes around $60\%$. This high similarity between items in a list may bore the user and negatively affect the overall user satisfaction. While a larger $p$ can help a little, P-LBRS fails to provide a flexible framework to balance the $\pazocal{R}$ and $\pazocal{D}$ trade-off. We solve this problem using H-LBRS.

\begin{figure}
\includegraphics[width=\linewidth]{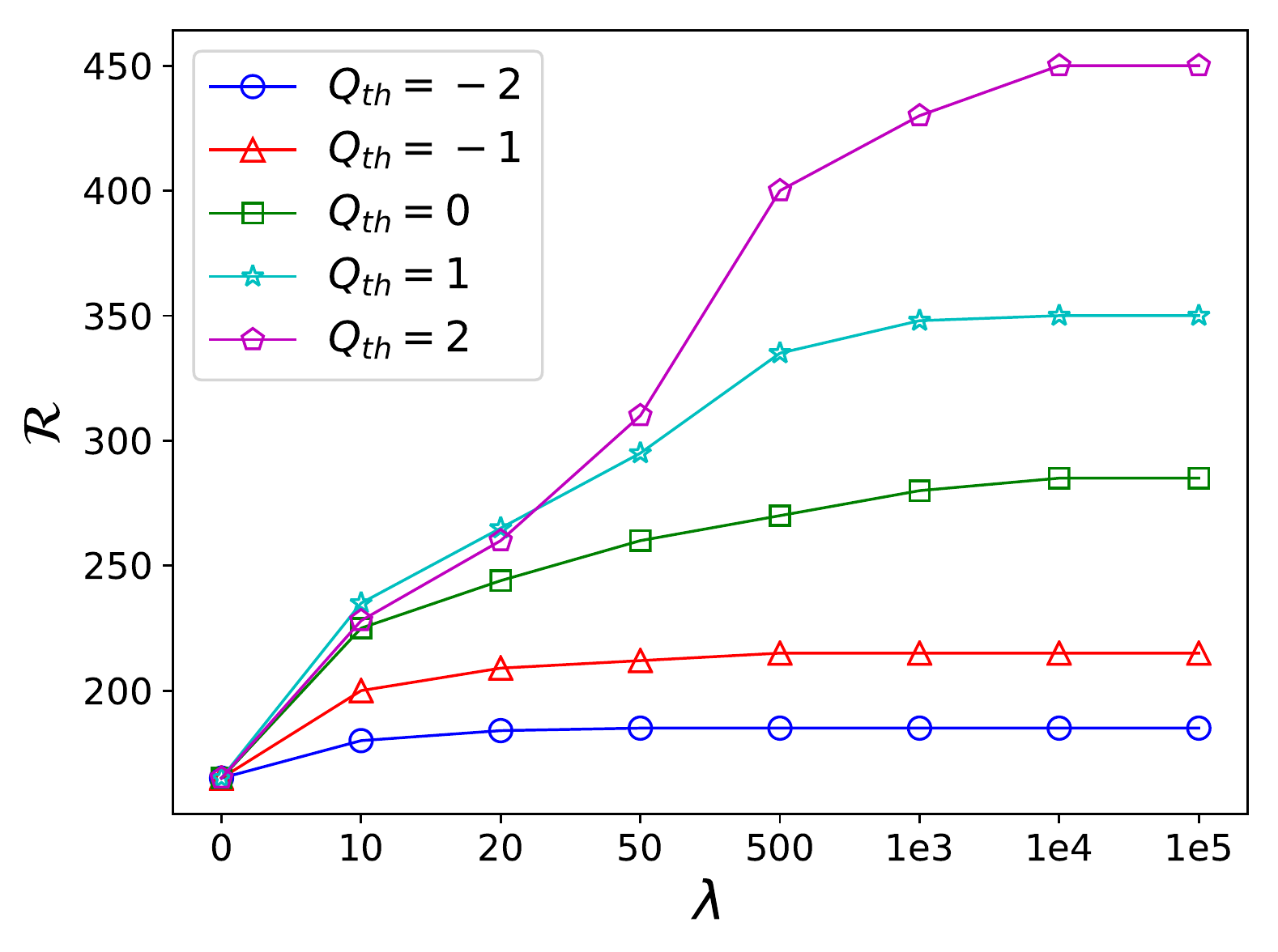}
\caption{The effect of $\lambda$ and $Q_{\mathrm{th}}$ on $\pazocal{R}$ in H-LBRS}
\label{fig:hlbrs-R}
\end{figure}

\begin{figure}
\includegraphics[width=\linewidth]{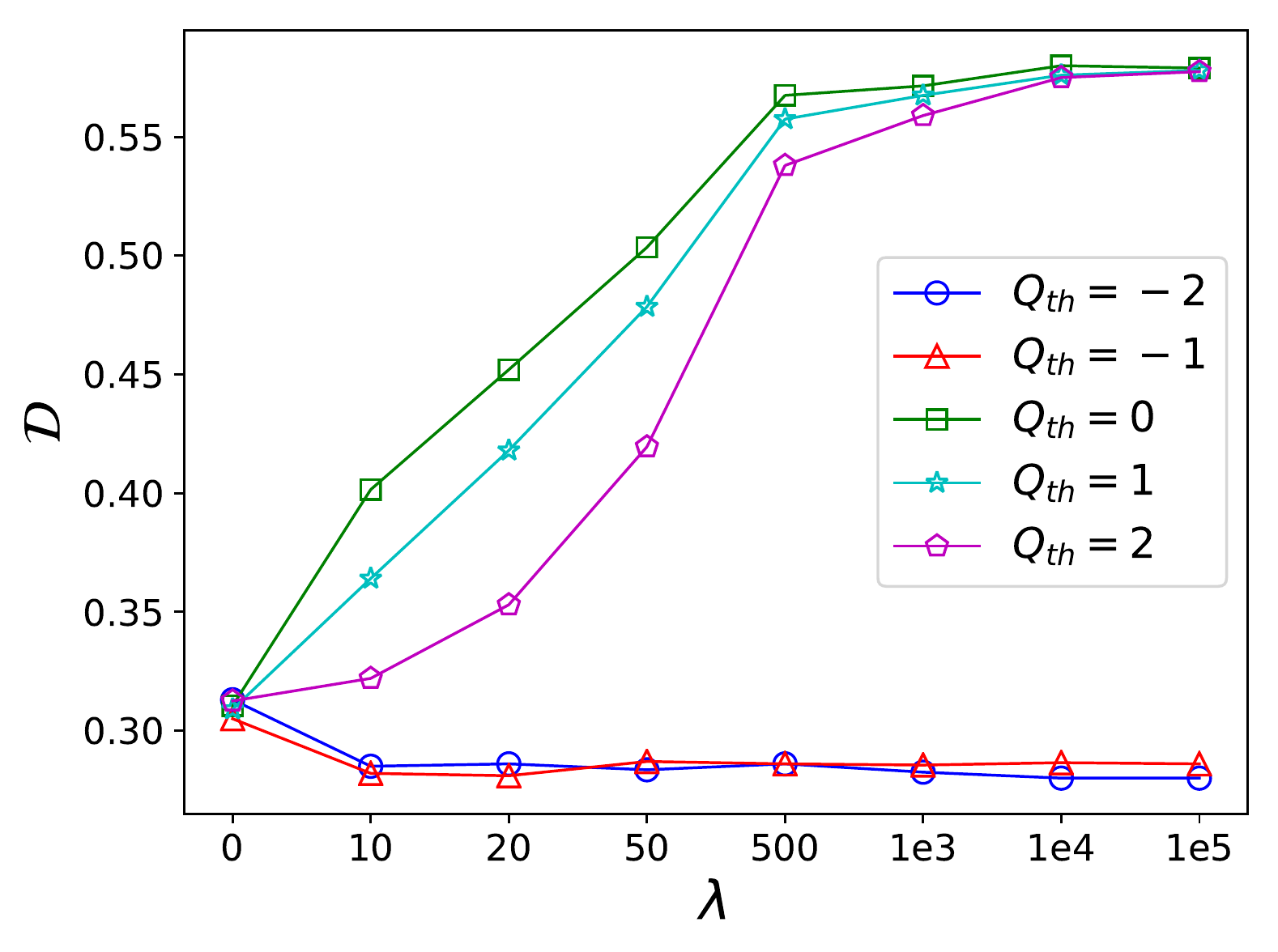}
\caption{The effect of $\lambda$ and $Q_{\mathrm{th}}$ on $\pazocal{D}$ in H-LBRS}
\label{fig:hlbrs-div}
\end{figure}

In H-LBRS, not only $p$ and $k$, but also $\lambda$ and $Q_{\mathrm{th}}$ affect the performance.  Fig.~\ref{fig:hlbrs-R} shows $\pazocal{R}$ for various values of $\lambda$ and $Q_{\mathrm{th}}$ when $k=5$ and $p=\frac{k}{100}$. We have considered extremes and varied $\lambda$ from 0 to 100,000, and also $Q_{\mathrm{th}}$ from -2 to 2. As it is seen, the increase in the value of both parameters can improve $\pazocal{R}$.  This increase is steady and converges at some point for different values of $Q_{\mathrm{th}}$. The least improvement is for $Q_{\mathrm{th}}=-2$, for which the increase in $\lambda$ slightly improves $\pazocal{R}$ and quickly becomes stable at 185 when $\lambda=20$. On the other hand, when $Q_{\mathrm{th}}=2$, $\pazocal{R}$ dramatically grows and becomes stable at 450 when $\lambda=10$,000, which improves $\pazocal{R}$ by 170$\%$ when $Q_{\mathrm{th}}=-2$.  However, this is only part of the story; diversity in H-LBRS is significantly decreased with the increase in the value of $\lambda$ and $Q_{\mathrm{th}}$, depicted in Fig.~\ref{fig:hlbrs-div}.  More specifically, although $\pazocal{D}$ is below 0.3 when $Q_{\mathrm{th}} \leq -1$, it reaches more than 0.55 when $Q_{\mathrm{th}}\geq 0$ and $\lambda\geq 500$. This clearly shows the trade-off between $\pazocal{R}$ and $\pazocal{D}$ discussed earlier.  From these results, one may ask which values are the best for $\lambda$ and $Q_{\mathrm{th}}$? While the response to this question generally depends on the designer and the application at hand, H-LBRS provides a flexible framework to meet desirable similarity and diversity constraints. To better demonstrate this ability of H-LBRS, we compare its performance with two values of $\lambda$ to that of other methods in the next section.
Moreover, we observed the same pattern as B-LBRS for H-LBRS for the effect of $p$ and $k$ on the performance metrics (i.e., while H-LBRS is insensitive to $p$, $\pazocal{R}$ and $\pazocal{D}$ moderately improve with a larger $k$), so we decided to not report repetitive results.

\subsubsection{Comparison with Baselines}
\label{subsub:comparison}

\begin{figure}
\includegraphics[width=\linewidth]{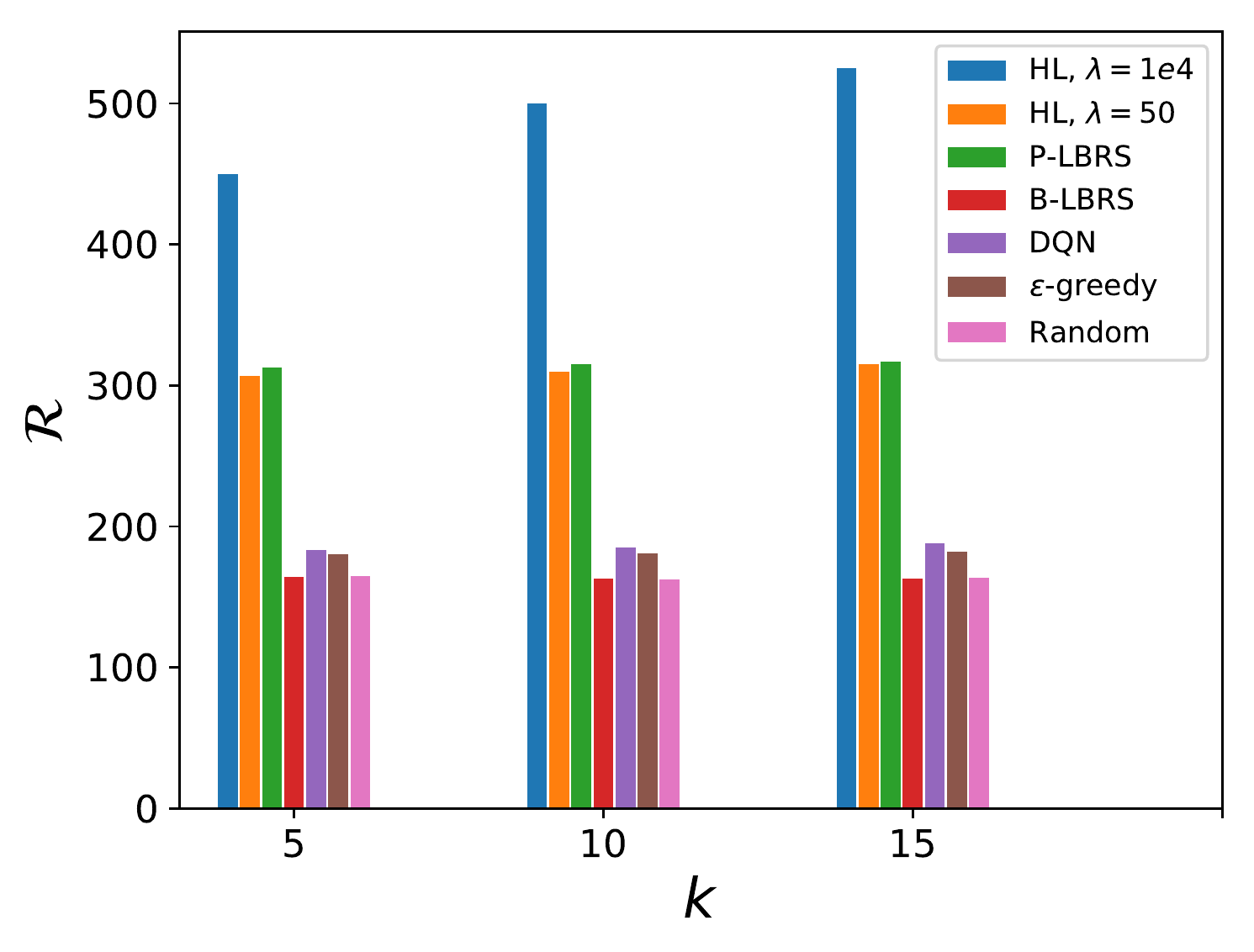}
\caption{Comparison between algorithms and the effect of $k$ on their performance}
\label{fig:varK}
\end{figure}

In this section, the results of experiments for comparison between methods are presented. As stated earlier, we include two versions of H-LBRS, indicated as HL in results, with $Q_{\mathrm{th}} =2$ and two values of $\lambda=50$ and 10,000. 
Fig.~\ref{fig:varK} illustrates $\pazocal{R}$ for all algorithms when $k$ grows from 5 to 15. Among the algorithms, H-LBRS with $\lambda=10$,000 significantly performs better than other methods. On average, it outperforms other methods between 40 to 170$\%$. The reason of this supremacy is that H-LBRS, with $\lambda=10$,000, recommends high quality documents--- those with $Q>2$--- more often. These documents not only replenish $B_t$, but also they can change user interests toward high quality topics. Apart from H-LBRS, P-LBRS performs the second best, with the same superiority reason.  DQN and $\epsilon$-greedy perform slightly better than B-LBRS and Random methods, as they have the learning ability. Both B-LBRS and Random methods achieve almost the same $\pazocal{R}$; it makes sense because both methods recommend documents randomly.  The last observation is that, as expected, all algorithms perform moderately better with a larger $k$. 

\begin{figure}
\includegraphics[width=\linewidth]{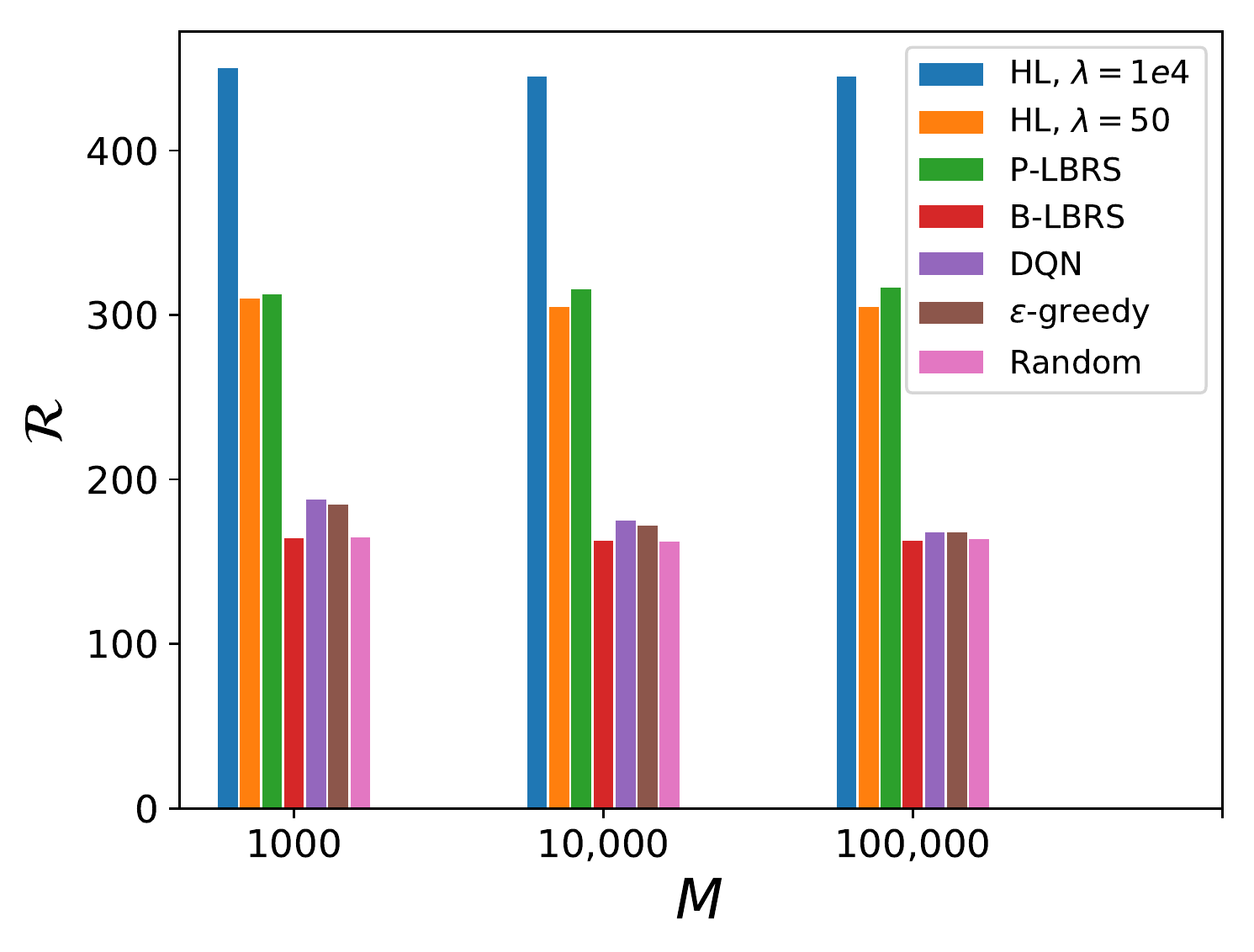}
\caption{Comparison between algorithms and the effect of $M$ on their performance}
\label{fig:varD}
\end{figure}

To see the effect of items scale, Fig.~\ref{fig:varD} depicts $\pazocal{R}$ for various values of $M$. The same pattern for algorithms' performance is seen here. While H-LBRS and P-LBRS considerably outperform other methods, their performance is not affected with a larger $M$. This obviously shows the robustness and scalability of these methods. Among the methods, DQN and $\epsilon$-greedy are negatively influenced the most by the increase in the value of $M$.  This is justified with the fact that as these methods are learning methods, when $M$ is too large, they can not learn the value of all items, performing almost similar to a random method.

\begin{figure}
\includegraphics[width=\linewidth]{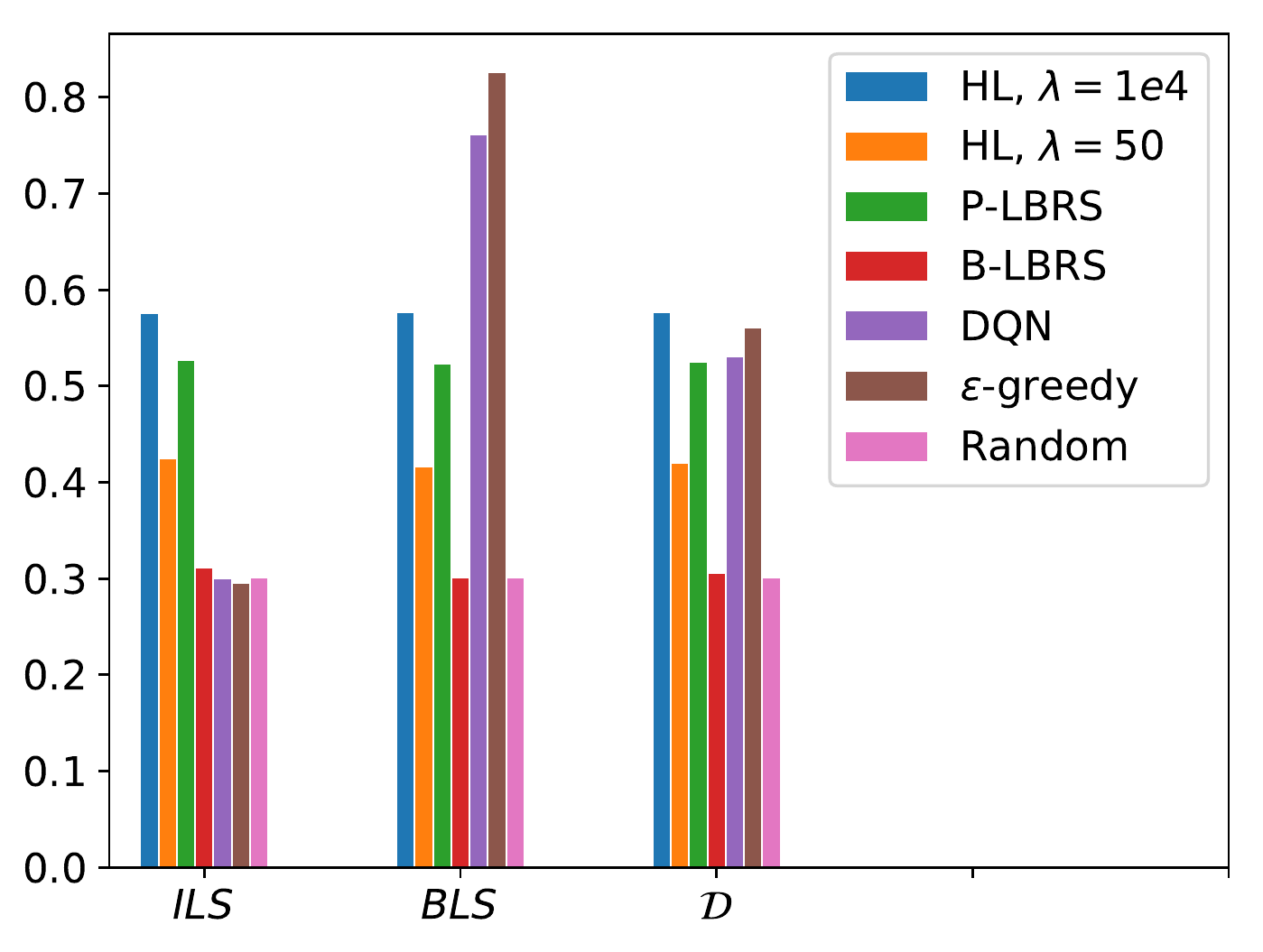}
\caption{Comparing diversity between algorithms}
\label{fig:all-div}
\end{figure}

Finally, Fig.~\ref{fig:all-div} illustrates diversity for all the methods. For the sake of completeness, we show $\pazocal{D}$, $ILS$, and $BLS$. Other than DQN and $\epsilon$-greedy, $ILS$ and $BLS$ are almost similar in all methods, leading to a similar $\pazocal{D}$ according to Eq.~\eqref{eq:div}.  On the other hand, $BLS$ is significantly larger than $ILS$ for DQN and $\epsilon$-greedy, leading to a large $\pazocal{D}$ that almost equals that of H-LBRS, $\lambda=10$,000 and P-LEACH. The reason of a large $BLS$ for DQN and $\epsilon$-greedy is that these methods perform mostly greedily and select the same set of documents to recommend to the users.  

\subsection{Discussion and Future Plan}
Several observations can be made from results presented in section~\ref{subsec:res}. First, while H-LBRS with $Q_{\mathrm{th}} =2$ and $\lambda=50$ can perform as good as P-LEACH in terms of $\pazocal{R}$, it improves diversity by about $25\%$. This can be concluded that one can tweak H-LBRS to effectively balance the similarity and diversity trade-off for a specific application, which can maximize the user satisfaction in the long term~\cite{bradley2001improving}. Second, random methods, including B-LBRS and Random, can yield the best diversity ($\pazocal{D}<0.3$, see Fig.~\ref{fig:all-div}). 
Third, while a method can have a reasonable intra-list diversity ($ILS$), it can perform quite poorly in terms of between-list diversity ($BLS$). This observation emphasizes the importance of taking into account $BLS$ in computing diversity in a sequential human-computer interaction. 
 
 Both P-LBRS and H-LBRS methods are built under the assumption that the RS knows the exact quality of items. This assumption, albeit strong, could be generally true in the real. For instance, human editors of a news agency can select high-quality articles for the \textit{featured} tab of their news platform~\cite{li2010contextual}.  This knowledge about the quality of items could be achieved by defining several quality attributes and scoring items based on these attributes. Or we might ask users to explicitly rank a list of quality attributes and then each item is matched against the user preferential ranking, receiving a quality score.
Another method to elicit the quality of items is to gradually learn it through interaction with the user~\cite{ie2019reinforcement}.  In general, it is more consistent with reality if the system can work with a more relaxed assumption about the quality of items, learned through one of the methods stated. We leave this for our future work.

 Finally, it is agreed that performance evaluation of RSs is cumbersome~\cite{ricci2011introduction}. This is mainly because, in order to evaluate the performance, the best way is to interact with a real user to have their feedback on recommendations made~\cite{beel2015comparison}. However, this process is very costly. A popular alternative is to use simulation as we used in this paper and is widely used by researchers in the RS field~\cite{zhao2017deep, zhao2018deep, zhao2018recommendations}.  Following our discussion above, if we determine the quality of items using one of the methods explained, we speculate that LBRS can be used and is effective in practice as well. This is also in our future work agenda; i.e., we plan to examine the performance of LBRS in an online study using real users.

\section{Conclusion}
\label{sec:con}
In this paper, inspired by LEACH protocol from WSNs literature,  we proposed a new recommendation approach, called LBRS, which is a simple, adaptive, scalable RS and can solve the cold-start and diversity problems in recommendation.  We presented three variants for LBRS; B-LBRS does not discriminates between items and uses a probabilistic approach for recommendation.  P-LBRS, on the other hand, factors in the importance of items and recommends items with higher qualities more often. Finally, H-LBRS considers a heterogeneity among items and provides a flexible structure to balance the similarity and diversity trade-off.  The results of experiments performed on RecSim validated the effectiveness of LBRS.  


\bibliographystyle{unsrt}
\bibliography{main}

\end{document}